\documentclass[twocolumn,showpacs,amsmath,amssymb]{revtex4}

\usepackage{graphicx}% Include figure files
\usepackage{dcolumn}% Align table columns on decimal point
\usepackage{bm}% bold math

\begin{document}

\title{Single-photon cooling in a wedge billiard}

\author{S. Choi$^{1}$, B. Sundaram$^{1}$, M. G. Raizen$^{2}$}
\affiliation{$^1$ Department of Physics, University of
Massachusetts, Boston, MA 02125, USA \\
$^2$ Department of Physics, The University of Texas, Austin, Texas
78712, USA}

\begin{abstract}
Single-Photon Cooling (SPC), noted for its potential as a versatile
method for cooling a variety of atomic species, has recently been
demonstrated experimentally. In this paper, we study possible ways
to improve the performance of SPC by applying it to atoms trapped
inside a wedge billiard. The main feature of  wedge billiard for
atoms, also experimentally realized recently, is that the nature of
atomic trajectories within it changes from stable periodic orbit to
random chaotic motion with the change in wedge angle. We find that a
high cooling efficiency is possible in this system with a relatively
weak dependence on the wedge angle, and that chaotic dynamics,
rather than regular orbit, is more desirable for enhancing the
performance of SPC.
\end{abstract}

\maketitle

\section{Introduction}

Single-Photon Cooling (SPC) is a general cooling method applicable
to most of the Periodic Table as well as molecules\cite{Price}.
Based on one-way wall of light\cite{OneWay1,OneWay2,OneWay3}, it
relies on irreversible optical pumping with one photon scattering to
trap atoms inside an optical dipole trap. The irreversibility is
achieved by exciting, with a depopulation beam, magnetically trapped
atoms to an intermediate hyperfine level from which they decay into
an optically trappable state  with a finite probability. This
process accumulates atoms inside an optical dipole trap without the
need for a cycling transition. The key mechanism behind SPC is that
only those atoms with kinetic energy less than a threshold energy
are captured by the optical trap i.e. cooling is achieved by
``filtering'' and then isolating colder atoms inside an optical
trap. The optical dipole trap is constructed in the form of a box in
SPC experiments and hence is simply referred to as the
``box''\cite{Price}.
%For the case of Rubidium atoms in a magnetic trap, irreversibility
%is achieved by exciting the magnetically trapped ($F= 2$) atoms to
%the $F'=1$ state by a depopulation beam which decays to the
%optically trappable $F=1$ state with a certain probability ($\sim
%84\%$ reported in the experiment).

The nature of SPC implies that a high cooling efficiency can be
achieved if as many atoms as possible  can be  made to encounter the
one-way wall of the box with the right kinetic energy. However,
various practical constraints exist to limit the cooling efficiency.
For instance, it is not productive to move the box around and stir
the atomic cloud with a depopulation beam as this can result in
significant loss of atoms -- there is a high probability that the
atoms encountering the box do not have the right energy to be
trapped. The box is therefore assumed to be stationary throughout
the evolution.  The finite size of the box means many of the atoms
simply will not pass through the position where the box is placed.
The usual isotropic trap is not necessarily the best potential for
SPC, since the atomic trajectories within an isotropic trap are
evenly spread rather than localized near a region where the box can
be placed for efficient collection of the atoms. The energy
consideration means that even if the trajectories of the atoms are
densely concentrated near one region it is of no use if the atoms
are moving too energetically to be trapped.

In order to find the best conditions for SPC, we consider in this
paper a special system in which various types of atomic trajectories
can be demonstrated: the wedge billiard for atoms. Wedge billiard
for atoms is a realization of the symmetric gravitational wedge
originally introduced by Letihet and Miller\cite{LM} which, owing to
the singularity of the vertex, demonstrates amazingly rich physics
--  the atomic trajectories can be tuned from stability to chaos
with the change in wedge angle. The wedge billiard for atoms has
already been implemented experimentally\cite{Milner} where the
atomic motion in different regimes of classical chaos were observed
and compared with numerical simulations. In this paper, we consider
SPC in wedge billiard and, more specifically, look for the best
position to place the box for all possible wedge angles. The best
box position is defined as the position at which the most number of
atoms is captured.

The overall goal of a cooling scheme is to increase the phase space
density. Phase space density in the context of cooling is defined as
the number of atoms in a box with sides of one thermal de Broglie
wavelength\cite{Townsend}. This can be written as $\rho = (N/V)
\lambda_{dB}^3$ where $N$ is the number of atoms, $V$ the volume,
and $\lambda_{dB}$ is the thermal de Broglie wavelength
$\lambda_{dB} = h/\sqrt{2 \pi m k_B T}$ where $h$ is the Planck
constant. This implies that the factor by which the phase space
density changes is given by
\begin{equation}
 \frac{\rho_f}{\rho_i}  = \frac{N_{box}}{N_{i}}
\frac{A_{wedge}(\theta)}{A_{box}} \left ( \frac{T_{i}}{T_{f}} \right
)^{3/2} , \label{ratio}
\end{equation}
where $A_{wedge}$ and $A_{box}$ are the areas of the 2-dimensional
wedge and the box, and $T_i$ and $T_f$ denote  respectively the
initial and final temperature. The fraction of atoms captured by the
box, $N_{box}/N_{i}$ is therefore an important measure, although it
has less impact than the ratio of the final to initial temperature
which is raised to the power of $3/2$. We shall define in this paper
the logarithm of Eq. (\ref{ratio}) as the cooling efficiency
\begin{equation}
\eta = \log_{10}(\rho_f / \rho_i) . \label{eta}
\end{equation}
We are most interested in how the atoms can be made to attain the
right energy and how the cooling efficiency is affected by the
regular and chaotic dynamics; from this, one may gain understanding
of how best to implement the SPC in general.

%To create an optical wedge billiard, two acousto-optic deflectors
%are employed, and by rapidly scanning the laser beam a time-averaged
%potential barrier of billiard walls can be generated. The atomic
%motion is confined to the 2-dimensional plane by adding a stationary
%standing wave along the optical axis -- the plane is then defined by
%the nodes of the standing wave.

The paper is organized as follows: In Section II we discuss how the
numerical simulation is carried out, including the parameters used
and the three types of boxes we use in our investigation. In
particular, we discuss how the results may be generalized, and
identify the core parameter space for this system, namely the wedge
angle, the initial temperature of the atoms and the box threshold
energy. In Section III, we present the numerical results for the
case that corresponds to the actual experimental parameters, and lay
the foundation for the subsequent section by showing how various
quantities such as the fraction of trapped atoms change for
different box types and wedge angles. In Section IV we present the
result of simulations that cover various combination of the three
core parameters to establish the general trend regarding the cooling
efficiency of SPC in a wedge billiard for atoms. We conclude in
Section V.

\section{Simulation Method}

\subsection{Atomic trajectories in wedge billiard}

First, we assume that atomic wedge billiard contains thermal atoms.
The initial positions and momenta of the atoms are assigned
according to a Gaussian random distribution as is usually done for
thermal atoms in equilibrium. In particular, the initial momenta of
the atoms are determined from the Maxwell-Boltzmann distribution
$f(v_x, v_y) = \left (\frac{m}{2 \pi k_{B}T} \right )^{3/2} \exp
\left [ - \frac{m (v_x^2 + v_y^2)}{2 k_{B}T} \right ]$ such that the
variance of velocity  $\sigma^2 = k_{B}T/m$ is proportional to the
the initial temperature of the atomic cloud, $T$. Here $k_B$ is the
Boltzmann constant and $m$ is the mass of the atom. Assuming hard
walls for the wedge billiard such that  atoms undergo elastic
collisions with the walls and taking the scattering cross section
for the interatomic collision to be zero, we calculate the expected
trajectory of each atom within the wedge billiard using classical
kinematic equations as done in Ref. \cite{Milner}. The calculated
trajectory for each of the atoms is stored and used later for the
analysis involving the box.

The number of atoms in the sample, $N$, is typically be of the order
$\sim 10^6$ experimentally but in order to simulate a realistic
system using a computationally manageable number of atoms we take
$N$ to be a few hundred and average the final result over several
runs with different initial position and velocity (but with the same
velocity variance i.e. the same initial temperature). With 200 atoms
in each run, averaging over 50 runs gave results similar to a single
run with 10000 atoms. For each run, the $N$ trajectories
corresponding to that particular set of initial conditions were used
to calculate the fraction of atoms trapped and lost by each type of
the box discussed below. These fractions are later averaged over the
number of runs to give a better estimate for a realistic atomic
sample. The averaging has the effect of smoothing out any large
fluctuations in the result; in fact, since we have a conservative,
closed system, the {\it fraction} of atoms captured by the box was
found to be fairly consistent for all $N$.

It was also found from our simulations that a significant proportion
of atoms  bounce off the walls and escape the wedge billiard,
especially in the chaotic regime; such atoms obviously cannot be
trapped by the box and are considered lost from the system. In an
attempt to avoid such loss and to perhaps improve the performance,
we have tried modulating the wedge angle during the time evolution
but found that it didn't lead to any better result. Such a set up is
also likely to be difficult to implement experimentally. We
therefore consider in our simulations wedge billiard with the walls
fixed at one wedge angle at a time throughout the entire duration of
the experiment.

\subsection{Modeling the box}

Numerically, the box is simply modeled as a region in space with a
certain implicit threshold energy $E_{box}$. The kinetic energy of
individual atoms varies greatly throughout the evolution, with the
majority of the atoms taking on kinetic energies far exceeding
$E_{box}$. For a typical $i$th atom with initial position $(x_i,
y_i)$, initial velocity $(v_x^{i}, v_y^{i})$,  and initial total
energy   $E_i = \frac{1}{2} m (v_x^{i 2} + v_{y}^{i 2})  + mg y_i$
where $g$ denotes gravitational acceleration, there are times
$t_{k}$, $k = 1,2,3 \dots$ during its evolution where the atom
reaches a certain height $y$ that overlaps with the position of the
box such that $E_i - mg y \leq E_{box}$. At these times, the atom
may be captured by the box. However not all atoms are captured in
reality: based on Ref. \cite{Price}, the efficiency of the box is
typically $\sim 85\%$ i.e. only $85\%$ of the atoms that pass
through the box ``sees'' the box (converted to the right hyperfine
state). This means $85\%$ of the atoms that pass through the box are
either lost (too high energy) or trapped (correct energy) and $15\%$
just passes through. We include this constraint in our simulations.
It is noted that since the atoms that pass through the box with
kinetic energy between zero and $E_{box}$ are assumed confined by
the box, the final equilibrium temperature of the atoms inside the
box can be estimated to be of the order $E_{box}/2k_B$. Finding the
right position to place a box given that there are $N$ atoms with
different trajectories is a particularly difficult optimization
problem, and is the main goal of this paper. To solve the problem of
where to place our box 3 types of boxes were considered in our
simulation: The optimum box and two types of fixed box we refer to
as Type I and Type II box. We explain these in more detail below.

\subsubsection{Optimum box}

Since we are considering a deterministic system, it is, in
principle, possible to calculate precisely where and when each  atom
reaches the required kinetic energy for it to be trapped. The
optimum box is a hypothetical box of vanishing size that, by
methodically moving around the trap, captures all theoretically
trappable atoms in sequence. This then gives the upper limit to the
number of atoms that can be captured for a given set of parameters.
To provide a more useful theoretical upper limit to the number of
trappable atoms, we additionally impose a couple of practical
constraints. One of these constraints is that the optimum box is
assumed not able to be at more than one position at one given
instant -- if there are more than one atom attaining trappable
energy at the same time only one of them is considered captured and
the rest are let evolving. The optimum box subsequently catches the
next available atom that reaches the correct energy, and so on. The
other constraint is, in line with the real experiment, we assume the
efficiency of the optimum box to be $85\%$. i.e. only $85\%$ of the
atoms passing through the optimum box are captured ({\it all} $85\%$
are captured since, by definition, all atoms intercept the optimum
box with the correct kinetic energy).

\subsubsection{Type I and Type II (fixed) box}

In reality, the box is finite in size and cannot be moved around
freely. We consider two possible cases in relation to where to place
the fixed box. One is to use the result of the optimum box
calculation above to guide us where to place our box. Based on the
optimum box calculation which gives a sequence of box positions over
the duration of the experiment, we choose to place a real,
finite-sized, stationary box at the position where the largest
number of the calculated trappable positions fall within the width
of box, in both $x$ and $y$ directions.  We call this stationary box
Type I box. On the other hand, in the absence of an optimum box
calculation, the best one can do is to actually place the finite
fixed box at various different places inside the wedge billiard to
find, by trial-and-error, the position where the most atoms can be
trapped. Given the symmetry of the wedge billiard system around
$x=0$, after collecting results corresponding to all possible
heights of the box on the axis of symmetry, we found the one height
$y$ that traps the most atoms for a given wedge angle. We refer to
this box obtained from ``optimization by hand'' as Type II box.

\subsection{Parameters used}

Using realistic parameters is obviously crucial for the correct
modeling of the system being simulated. With too big or too small
dimensions, the atomic trajectories and hence the calculated cooling
efficiency is likely to be unrealistic. To address this issue we
base our parameters on the experimental values\cite{Price,Milner}
and choose our parameters to be within reasonable range of these
values. In addition to simulating a realistic system, one should
ideally be able to generalize the result beyond the existing
experiment. However it is noted that, especially in the chaotic
regime, it is impossible to write down analytically a general
expression of the atomic trajectories as a function of various
parameters. This limits our options to only those of numerical
analysis. Since one cannot cover every single possible value of
various parameters, only a general trend can be identified from a
numerical study. Such general trend should, however, provide
sufficient information to understand the fundamental physics of the
system, and give us a clue as to the range of realistic cooling
efficiencies possible with this system.

The interdependence of various parameters means  the parameter space
boils down to three independent variables: the wedge angle,  initial
temperature, and the box threshold energy. It is noted first of all
that the regular and chaotic behavior exhibited by the atoms inside
the wedge billiard is dependent only on the wedge angle, and is
independent of factors such as the atomic species, initial
temperature, and the qualities of the box. We therefore capture all
the necessary physics in our simulation by scanning through all
possible wedge angles. Also, in our numerical model, different
atomic species is represented only via their different atomic mass
which shows up in the velocity variance $\sigma^2$ and in the
calculation of the kinetic energy $E_k = \frac{1}{2}m v^2$. It is
noted that any difference in physics due to different atomic mass is
taken care of with an appropriate change in the initial temperature
$T_i$ ($\sigma^2 \propto T_i$), and the kinetic energy $E_k$ is
automatically scaled accordingly. This can be shown by setting $k_B
= T_i = m = 1$ so that the initial thermal energy $E = k_B T_i = 1$
becomes unit energy and $\sigma = \sqrt{k_{B}T_i/m} = 1$ becomes
unit velocity such that the unit length $l$ can be defined as $l =
\sigma \tau$ where $\tau$ is unit time. The kinetic energy of an
atom traveling with velocity $v$ is then given  in the scaled unit
as $E_{k} = \frac{1}{2} (v/\sigma)^2$ i.e. the kinetic energy
relative to  the initial thermal energy characterized by $T_i$ is
what matters. All the major physics of this system can therefore be
covered by studying various combination of wedge angle, initial
temperatures and the box threshold energies.

\begin{figure*}
\begin{center}
\includegraphics[height=9cm]{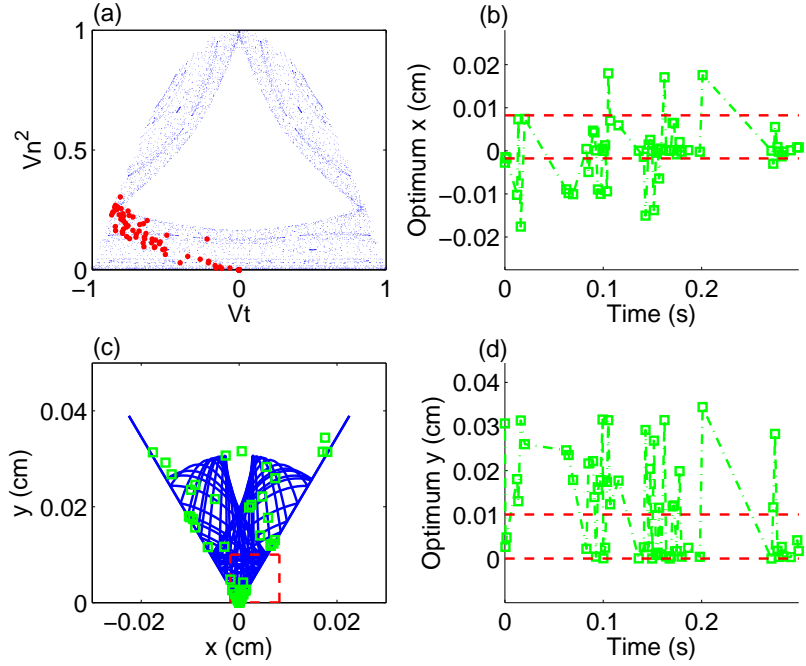}
\caption{(Color online) (a) Poincar\'{e} section, $v_t$ vs. $v_n^2$
for the wedge billiard with wedge angle $\theta = 30^\circ$ (smaller
blue dots). The larger red dots represent the distribution of the
thermal atoms after their first bounce on the wedge walls. The right
hand column, i.e. panels (b) and (d), show the $x$ and $y$ component
of the optimum box position over time as green squares. The dash-dot
line connects the green squares in sequence to show how the optimum
box position changes over time. The red dashed lines mark the edges
of the Type I box. (c) The wedge in real space with one typical
trajectory for an atom shown as blue line. The Type I box used in
our simulation is shown in the same panel as a red box in dashed
line. The green boxes mark the (changing) positions of the optimum
box over the duration of the simulation.} \label{Fig1}
 \end{center}
\end{figure*}

\begin{figure*}
\begin{center}
\includegraphics[height=9cm]{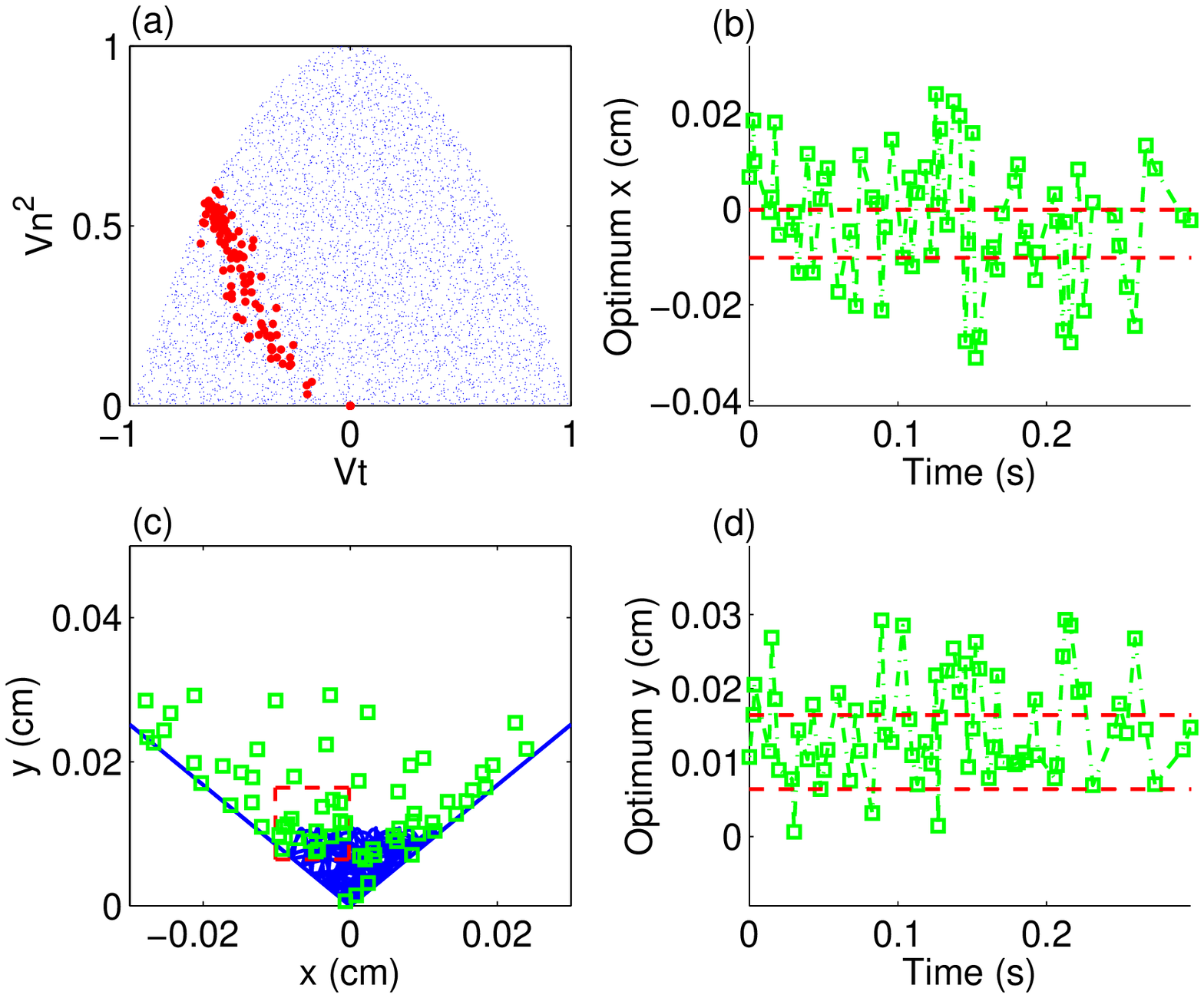}
\caption{(Color online) Same as Fig. \ref{Fig1} except the wedge
angle is now $50^\circ$.} \label{Fig2}
 \end{center}
\end{figure*}

\begin{figure*}
\begin{center}
\includegraphics[height=9cm]{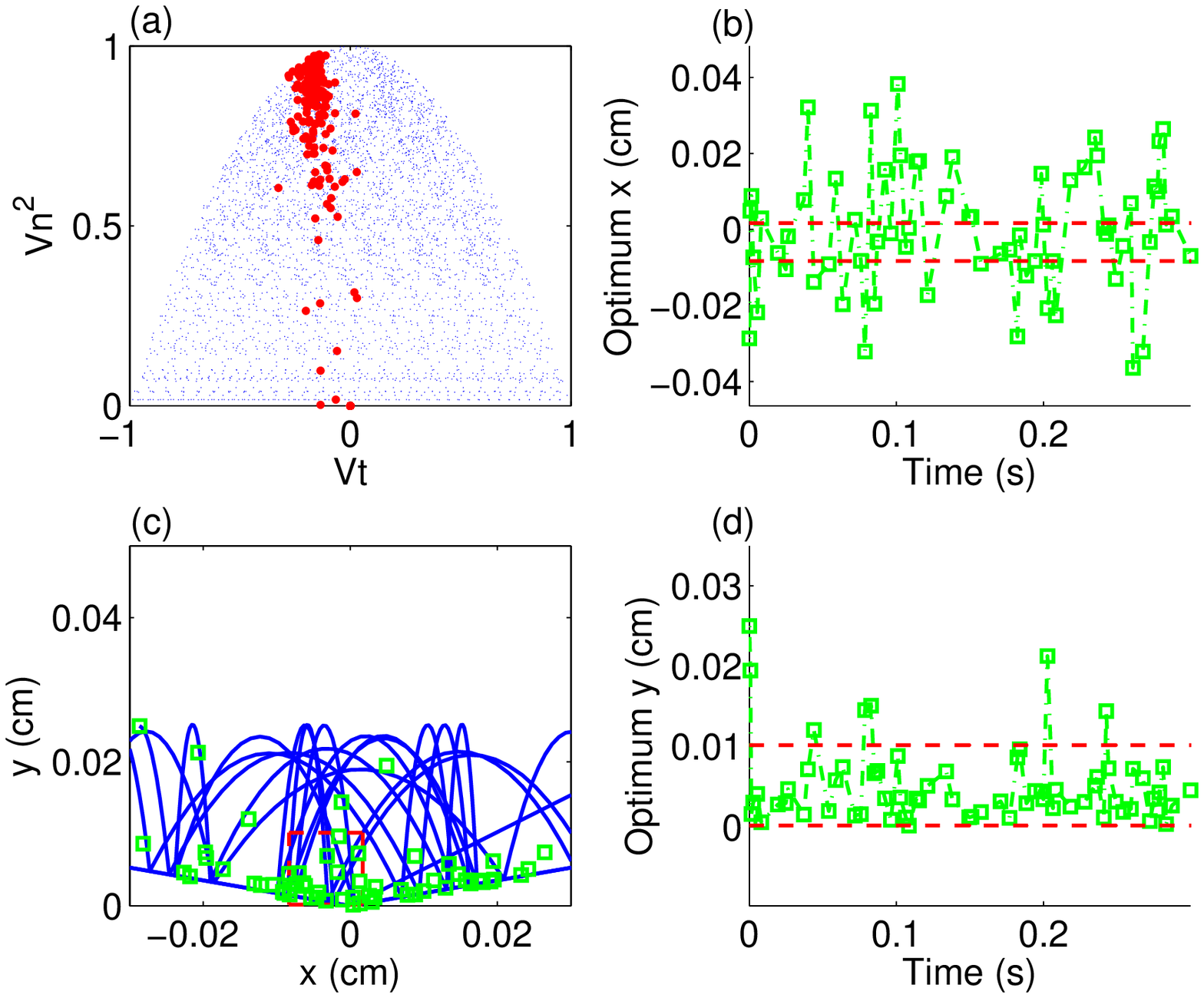}
\caption{(Color online) Same as Fig. \ref{Fig1} except the wedge
angle is now $80^\circ$.} \label{Fig3}
 \end{center}
\end{figure*}

\section{Results for several wedge angles}

%and a Gaussian probability density with the following mean and
%standard deviation in positions: $\bar{x} = 0 \/ cm$, $\bar{y} =
%0.01 \/ cm$, $\sigma_x = \sigma_y = 200 \times 10^{-4} \/ cm$,   and
%velocities: $\bar{v}_{x} = \bar{v}_{y}  =  0 \/ cm s^{-1}$,
%$\sigma_{v_x} = \sigma_{v_y}  = 0.75 \times 0.35 \/ cm s^{-1}$.
%The fact that we used 200 atoms and averaged
%the result over several runs shouldn't stop us from extrapolating
%the results to a more realistic number of atoms since from the phase
%space ($v_t$ vs. $v_n^2$) plots, it was found that even without
%averaging over many runs, 200 atoms was found to cover the phase
%space evenly within a short time.
In this section, we illustrate the physics of the system by
presenting the results of numerical simulation for select wedge
angles using the experimentally verified parameters of Refs.
\cite{Price} and \cite{Milner}. The results of our simulation for
various regimes characterized by the three different wedge angles
$30^\circ$, $50^\circ$, and $80^\circ$ are presented in Figs.
\ref{Fig1}-\ref{Fig3} respectively. In each of these figures, we
present several sub-figures. First of all, as one of the
sub-figures, we present the Poincar\'{e} section for the wedge
billiard typically presented as a plot of $v_t$ vs. $v_n^2$ where
$v_t$ and $v_n$ denote transverse and normal velocity of the atoms
immediately after hitting the walls. As is well known, visible
geometric structures or ``islands'' within the Poincar\'{e} section
represent regions of regular orbit while the regions containing
evenly spread dots correspond to chaotic dynamics. The fact that
there are no islands for wedge angles $50^\circ$ and $80^\circ$
(Figs. \ref{Fig2} and \ref{Fig3}) means that  all atoms are expected
to undergo chaotic dynamics, while the existence of an island for
wedge angle $30^\circ$ (Fig. \ref{Fig1}) means atoms with initial
conditions that lie in the island undergo regular orbits. The
initial thermal velocity distribution in terms of $v_t$ and $v_n^2$
after one bounce off the walls are shown as (thicker) red dots on
the Poincar\'{e} section as a guide to the initial condition used.
Next, typical trajectory of one particular atom in the wedge
billiard is presented as well as all the positions of the optimum
box in small green squares. The outline of the Type I box determined
from the optimum box calculation is also shown superposed in the
same sub-figure. In the right hand column of the figures, the
time-varying positions of the optimum box decomposed into $x$ and
$y$ components are presented along with the boundaries of the Type I
box shown in dashed lines.

\begin{figure*}
\begin{center}
\includegraphics[height=8cm]{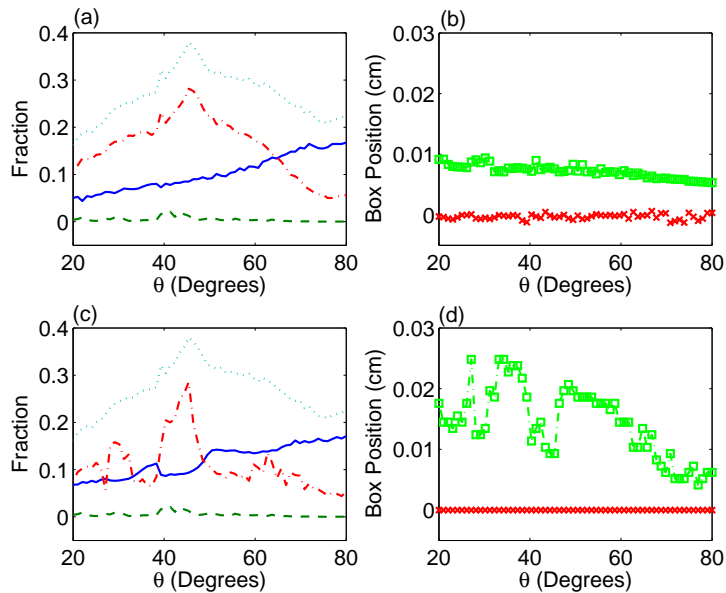}
\caption{(Color online) Panel (a) displays the final (after time $t
= 300ms$) fraction of trapped atoms (Solid line), atoms remaining in
the wedge (Dashed line), and atoms lost (Dash-dot line) for various
wedge angles $\theta \in [20^\circ, 80^\circ]$ for Type I box.  The
additional dotted line represent the fraction of trapped atoms for
the case of optimum box, i.e. the theoretical upper limit for
comparison  (b) The Type I box position for different wedge angles
-- the squares represent the $y$ component of the mean position and
the crosses mark the $x$ component. (c) and (d): Same as (a) and (b)
but for Type II box.} \label{Allangle}
 \end{center}
\end{figure*}

In Fig. \ref{Allangle} we present various fractions  as a function
of wedge angles -- atoms trapped by the box, atoms remaining in the
wedge, and atoms lost for all 3 box types. As mentioned previously,
there are two mechanisms by which the atoms are lost -- one by
encountering the box with the wrong energy and the other by escaping
the wedge billiard after the collision with one of the walls. We
also show the $x$ and $y$ components of the Type I and Type II boxes
as a function of wedge angle. It is found that the Type II box gives
slightly better result in terms of the fraction of atoms trapped. It
is seen that, for the parameters used, from $5\%$ to up to around
$15\%$ of the atoms could be trapped by this scheme. The fraction of
atoms trapped has (discounting fluctuations) almost linear
dependence on the wedge angle with the largest fraction of atoms
trapped at $\theta = 80^{\circ}$.  These observations imply that
chaotic dynamics, rather than the regular dynamics, is more
conducive for SPC. This makes sense since, to be trapped by the box,
the atom has to reach the correct height relative to the initial
energy that corresponds to the right kinetic energy.  This is more
likely in the regime of chaotic dynamics in which the atoms take on
various different trajectories over time rather than that of regular
orbit which generally has higher kinetic energy  and a limited range
of trajectories. The wide range of trajectories also means that more
atoms are likely to encounter the stationary box. The higher
fraction of atoms trapped for larger wedge angle can be explained
from the fact that with wider wedge angles, the normal velocity
component $v_n$ dominates that of the transverse velocity component
$v_t$ and hence one has a higher fraction of atoms in parabolic
motion. With parabolic motion it is easier for the atoms to attain
the trappable energy as the kinetic energy goes right down to zero
at the turning points.

Finally we present here a simple analysis regarding the height of
the best fixed box by assuming that the box is able to catch the
atom after one bounce from the walls. Taking the cusp of the wedge
billiard as the origin of position, and considering an atom
initially at position $x_i$ and $y_i$ with initial velocity
components $v_{x}^{i}$ and $v_{y}^{i}$, the $x$ and $y$ components
of the velocity on impact with the wall at angle $\theta$ from the
perpendicular line at the origin are $v_{x}^{w} = v_{x}^{i}$ and
$v_{y}^{w} = \sqrt{v_{y}^{i \; 2} - 2 g \left [ |y_i| - |x_i| \tan
\left (\frac{\pi}{2} - \theta \right ) \right ]}$. Energy
conservation implies:
\begin{equation}
mgh = \frac{1}{2} m \left ( v_{x}^{w \; 2} + v_{y}^{w \; 2}  \right
)  - E_{box} ,
\end{equation}
where $E_{box}$ is the threshold energy of the box. This gives the
best box height as
\begin{eqnarray}
h & = & \frac{E^{i}_K - E_{box}} {mg}  -   |y_i|  +  |x_i| \tan
\left
(\frac{\pi}{2} - \theta \right ) \label{hbest} \\
& \approx & \frac{E^{i}_K - E_{box}} {mg}  -   |y_i|  +  |x_i| \left
( \frac{1}{\theta} - \frac{\theta}{3} -  \cdots \right ) .
\end{eqnarray}
This result is valid for larger $\theta$ where the first bounce
results in a parabolic trajectory e.g. when $y_0 >  L_{w} \cos
\theta$ where $L_w$ is the length of the wedge wall. For smaller
$\theta < 50 ^{\circ}$ a regular orbit is likely which means any
analysis based on one bounce does not hold. A plot of $h$ given by
Eq. (\ref{hbest}) was found to closely reproduce the large angle
$\theta
> 50 ^{\circ}$ part of Fig. \ref{Allangle}(d).

The cases where $v_x$ is large enough (or $\theta$ is small enough)
so that the atom hits the wall due to the horizontal rather than the
vertical component of motion require a modified analysis: Assuming
the time it takes for an atom to hit the wall is $\tau_x$, the $y$
component of the velocity at the wall is  $v_{y}^{w} = v_{y}^{i} -
g\tau_x$ while $x$ component of the velocity at the wall is
$v_{x}^{i}$.  The vertical displacement after  $\tau_x$ is $y_i -
\frac{1}{2}g \tau_x^2$ and this means the distance traversed
horizontally  $ v_x^{i}\tau_x = \left [ y_i - \frac{1}{2}g \tau_x^2
\right ] \tan \theta $, i.e. $g \tau_x = \cot \theta \left [ -
v_x^{i} + \sqrt{v_x^{i \; 2} + 2 g y_i \tan^{2} \theta} \right ]$.
The energy conservation then implies
\begin{eqnarray}
h  & = & \frac{E^{i}_K - E_{box}} {mg}  -   v_y^{i}\tau_x +
\frac{1}{2}g \tau_x^2  \nonumber \\
& = & \frac{E^{i}_K - E_{box}} {mg}  +  y_i  -  \frac{\cot
\theta}{g} ( v_y^{i} +  v_x^{i} \cot \theta ) \nonumber \\
& &  \times  \left [ \sqrt{v_x^{i \; 2} + 2 g y_i \tan^{2} \theta} -
v_x^{i}\right ]   \nonumber \\
& \approx & \frac{E^{i}_K - E_{box}} {mg}  - \frac{y_i v_y^i}{v_x^i}
\theta
 + \frac{g y_i^2}{2  v_x^{i \; 2}}  \theta^2 + \cdots
 \label{hbest2}
\end{eqnarray}
This reproduces the result for smaller $\theta$ but again, since we
are only considering one bounce and ignoring the possibility of a
regular orbit, the oscillatory behavior is not recovered in this
very approximate result.

\section{Cooling Efficiency of SPC in Wedge Billiard}

In this section, we calculate the cooling efficiency $\eta$ for SPC
in wedge billiard.  In particular, we identify general features of
this system by considering various combination of the three core
parameters: wedge angle, initial temperature, and the box threshold
energy. To give a better sense of the range of parameter values used
in this section, we denote the initial temperature and the box
threshold energy of existing experiments as $T_{i}^{ex}$ and
$T_{b}^{ex}$ respectively and refer to all the other initial
temperatures and threshold energies as some multiples of
$T_{i}^{ex}$ and $T_{b}^{ex}$.

\begin{figure*}
\begin{center}
\includegraphics[height=8cm]{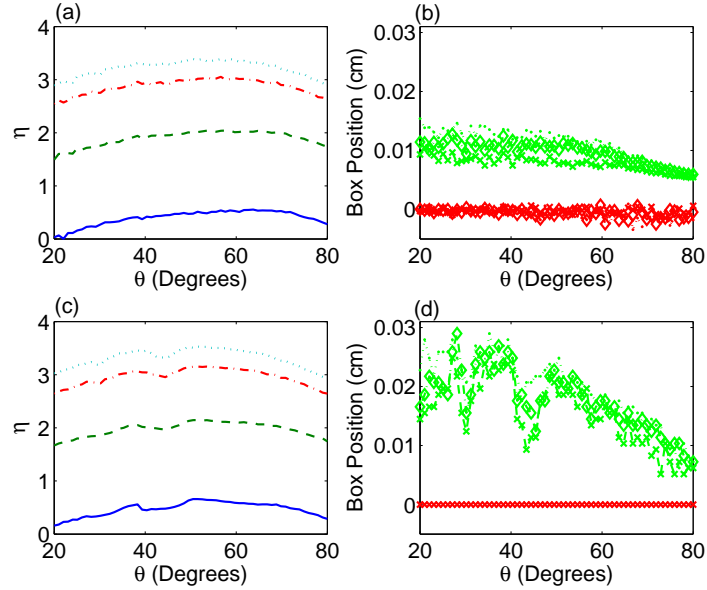}
\caption{(Color online) Panel (a) displays the cooling efficiency
$\eta = \log_{10} (\rho_f/\rho_i)$   for different initial
temperatures of $T_{i}^{ex}$ (Solid line), $10T_{i}^{ex}$ (Dashed
line), $50T_{i}^{ex}$ (Dash-dot line) and $100T_{i}^{ex}$ (Dotted
line) as a function of wedge angles  $\theta \in [20^\circ,
80^\circ]$ for Type I box. The box threshold energy is $T_{b}^{ex}$.
(b) Type I box position for different wedge angles -- the crosses,
diamonds, and dots represent the $y$ component of the mean position
for the initial temperatures $10T_{i}^{ex}$ (Dashed line),
$50T_{i}^{ex}$ (Dash-dot line) and $100T_{i}^{ex}$ (Dotted line)
respectively (Green in color in the online version). The
corresponding symbols around zero mark the $x$ component. (Red in
color in the online version.) (c) and (d): Same as (a) and (b) but
for Type II box.} \label{PSDInitTemp}
 \end{center}
\end{figure*}

\subsection{Changing the initial temperature}

In Fig. \ref{PSDInitTemp}, we present the cooling efficiency for
Type I and Type II boxes.  Figure \ref{PSDInitTemp} is arranged
similarly to Fig. \ref{Allangle}, except that we present the cooling
efficiency $\eta$ for the four different initial temperatures of
$T_{i}^{ex}$, $10T_{i}^{ex}$, $50T_{i}^{ex}$, and $100T_{i}^{ex}$
for the box threshold temperature of $T_{b}^{ex}$. We also show, as
done in Fig. \ref{Allangle}, the positions of Type I and Type II
boxes  as a function of wedge angle for different initial
temperatures. We chose to look at such widely varying initial
temperatures up to $100T_{i}^{ex}$ since it was found numerically
that smaller changes did not noticeably affect the cooling
efficiency e.g. the difference in result between say $T_{i}^{ex}$
and even $4T_{i}^{ex}$ was not noticeable. Such immunity to
temperature changes means the results shown here for each of these
four initial temperatures should hold for a wide range of atomic
species and length scales. It is seen that there is, in fact,  quite
a dramatic increase in phase space density -- the higher the initial
temperature the more significant is the improvement.

It is also seen that due to the interplay of various factors in the
calculation of $\eta$, the curves do not show very strong dependence
on the wedge angle, although one can notice slight changes in the
(shallow) maximum of the curves. The curves have maximum near
$70^\circ$ for the initial temperature of $T_{i}^{ex}$ and near
$50^\circ$ for the initial temperature of $100T_{i}^{ex}$. Given
that the ratio of the area $A_{wedge}(\theta)/A_{box}$ has maximum
at the wedge angle $\theta = 45^{\circ}$ this implies that the
fraction of atoms trapped (i.e. instead of $\eta$  that also
compares the temperature difference) does not change much with the
change in the wedge angle for higher initial temperature. Indeed,
the fraction of atoms trapped was found to increase almost linearly
with the increasing wedge angle, but the gradient of this linear
variation was highest for the initial temperature of $T_{i}^{ex}$
and the smallest for the initial temperature of $100T_{i}^{ex}$. The
fact that the fraction of atoms trapped was the highest for $\theta
= 80^\circ$ for all initial temperatures (albeit with a varying
degree) re-confirms the main mechanism by which the atoms get
trapped by the box -- with wider wedge angle, one has a higher
fraction of atoms in parabolic trajectory.

The change in phase space density corresponding to the change in
initial temperature from $50T_{i}^{ex}$ to $100T_{i}^{ex}$ is
smaller compared to that with change in initial temperature from
$10T_{i}^{ex}$ to $50T_{i}^{ex}$ i.e. there is a saturation in the
amount of increase in phase space density. This can be partly
explained by examining the fraction of atoms trapped by the box. It
was found that the fraction of atoms trapped did not change much at
all between the initial temperatures of $T_{i}^{ex}$ and
$10T_{i}^{ex}$ i.e. the temperature ratio has a major effect. Over
$10T_{i}^{ex}$, however, the fraction of trapped atoms began to be
noticeably reduced. This is because after the temperature gets high
enough atoms have enough energy to escape the wedge in significant
numbers. With many atoms escaping and reducing the pool of atoms
inside the wedge, the actual fraction of atoms trapped by the box
goes down, offsetting the effect of the larger temperature ratio.

Finally we note that the results of Type I and Type II are very
similar. This is because of two reasons -- one is that the maximum
fraction of atoms trapped using Type II boxes, while greater than
that using Type I box, is not significantly different. The second
reason is that the finite box size implies there is actually a
significant overlap between the two different types of box
positions.  The finite box size also means that the slightly
different box positions for different initial temperature (for both
Type I and Type II) is not noticeable -- the temperature-dependent
difference in the position of the box center is smaller than the
width of the box itself.

\begin{figure*}
\begin{center}
\includegraphics[height=8cm]{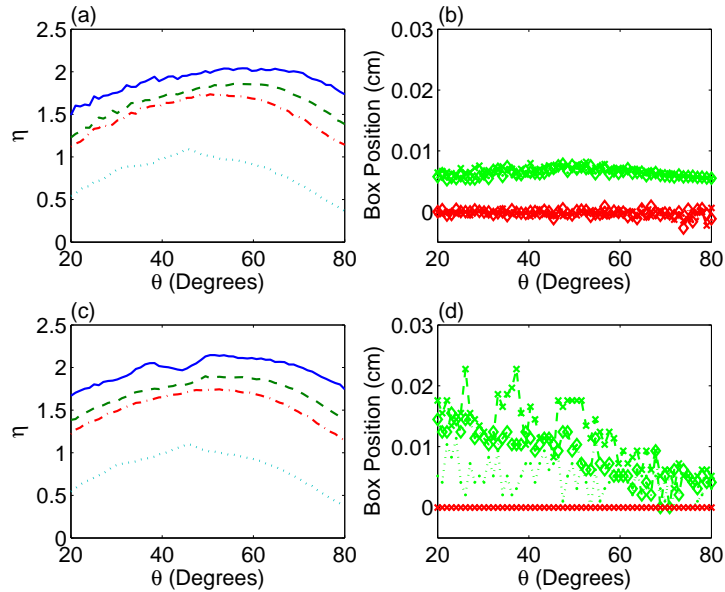}
\caption{(Color online) Panel (a) displays the cooling efficiency
$\eta = \log_{10} (\rho_f/\rho_i)$  for different box threshold
temperatures of $T_{b}^{ex}$ (Solid line), $2T_{b}^{ex}$ (Dashed
line), $3T_{b}^{ex}$ (Dash-dot line) and $10T_{b}^{ex}$ (Dotted
line) a function of wedge angle $\theta \in [20^\circ, 80^\circ]$
for Type I box. The initial temperature was $10T_{i}^{ex}$. (b) Type
I box position for different wedge angles -- the crosses, diamonds,
and dots represent the $y$ component of the mean position for the
box threshold temperatures $2T_{b}^{ex}$ (Dashed line),
$3T_{b}^{ex}$ (Dash-dot line) and $10T_{b}^{ex}$ (Dotted line)
respectively (Green in color in the online version). The
corresponding symbols around zero mark the $x$ component. (Red in
color in the online version.) (c) and (d): Same as (a) and (b) but
for Type II box.} \label{PSDBoxTemp}
 \end{center}
\end{figure*}

\subsection{Changing the box threshold energy}

Similarly to Fig. \ref{PSDInitTemp}, we show in Fig.
\ref{PSDBoxTemp} the cooling efficiency $\eta$ and the corresponding
positions for Type I and Type II boxes. But this time, we fix the
initial temperature at $10 T_{i}^{ex}$ and show the general trend
with respect to four different box threshold temperatures of
$T_{b}^{ex}$, $2T_{b}^{ex}$, $3T_{b}^{ex}$, and $10T_{b}^{ex}$.
 The initial temperature of $10T_{i}^{ex}$ was chosen to make sure that even with
box threshold temperature of $10T_{b}^{ex}$ the atomic sample is
actually being cooled down. The observed trend with increasing box
threshold energies is the opposite of the trend observed above with
increasing temperature: $\eta$ {\it decreases} with increasing box
threshold energy, which is unexpected since with increasing box
threshold energy, the actual fraction of atoms trapped is increased
as the atoms are more likely to be trapped. This can be understood
from the fact that, with increasing box threshold energy, the ratio
of initial to final temperature is decreased and the temperature
ratio is the more significant contributor to $\eta$.

Again, there is a relatively weak dependence of $\eta$ on the wedge
angle. The wedge angle at which the maximum $\eta$ occurs is
observed to be slightly shifted from near $70^\circ$ for the box
threshold energy of $T_{b}^{ex}$ to near $50^\circ$ for the box
threshold energy of $10T_{b}^{ex}$.  As before, this can be
explained by the considering the ratio of the areas
$A_{wedge}(\theta)/A_{box}$ and the fraction of atoms trapped. In
fact, the fraction of the atoms trapped as a function of wedge angle
demonstrates quite a different behavior from that seen above for
different initial temperatures. Although the fraction of atoms
trapped with box threshold energy of $T_{b}^{ex}$ has roughly linear
dependence on the wedge angle (with maximum near $80^\circ$), with
higher box threshold energies, the fraction of atoms trapped
gradually demonstrates a ``triangular'' shape as a function of wedge
angle (with maximum near $50^\circ$). Interestingly, with the box
threshold energy of $10T_{b}^{ex}$, it matches the triangular shape
seen for the optimum box result as dotted line in Fig.
\ref{Allangle}(a) and (c). This shows that with large enough box
threshold energy, all atoms that can be trapped as calculated by the
optimum box are indeed trapped given enough time. This is a
significant result since it shows a potential to ``simulate'' the
complicated optimum box calculation experimentally.

A saturation is found to occur as the box threshold energy becomes
larger; changes in $\eta$ due to changes in box threshold energy
become smaller as the box threshold energy becomes close to
$10T_{b}^{ex}$. This is to be expected since, once the box threshold
energy becomes large enough to catch all the atoms that are
physically feasible to be caught (i.e. ones that do not escape the
wedge billiard), any higher box threshold energy will not give
different fraction of atoms trapped.  As regards to the actual box
position itself and the use of Type I and Type II boxes, similar
observation  as above is made  i.e. the final cooling efficiency for
the two different box position is quite similar and that,  due to
the finite width of the box, difference in box position do not lead
to noticeable changes in $\eta$.

\section{Conclusion}

We have found that SPC  can significantly increase the phase space
density of the atoms originally trapped inside a wedge billiard. It
was found that even in the very tough scenario of very high initial
temperature and very low box threshold energy, enough atoms are
expected to be caught to give a high cooling efficiency. The cooling
efficiency $\eta$ showed a relatively weak dependence on the wedge
angle. This is because $\eta$ depends on the interplay of various
factors, not just the number of atoms trapped by the box. For low
initial temperature, and also for low box threshold energy, the best
angle was near $\theta = 70^\circ$ and for higher initial
temperature and also for higher box threshold energy,  the best
angle was near $\theta = 50^{\circ}$. Various trends could be
explained by studying the fraction of atoms trapped as a function of
the wedge angle. Both Type I and Type II boxes were found to give
similar cooling efficiencies, although the Type II box is one that
can be obtained experimentally in the absence of any knowledge about
the idealized optimum box calculation.

The regime in which the performance of SPC in wedge billiard
 is best was identified to be the regime of chaotic dynamics with
wedge angle $\theta \geq 45^\circ$. In
the chaotic regime, atoms take on various different kinetic energies
and trajectories, increasing the likelihood of meeting the condition
for SPC. Although on first sight atoms in regular orbits look more
promising, their kinetic energy is less likely to be redistributed,
resulting in a smaller cooling efficiency. Within the chaotic
regime, it was found that the wider wedge angle is better in terms
of capturing a higher fraction of atoms. This is because, with a
wider wedge angle, there are more of atoms undergoing parabolic
motion where the kinetic energy becomes zero at the turning point.
These observations indicate that, in general, the best set up for
SPC is the one where the box is able to access the most number
of turning points concentrated within a small region.  Somewhat less ideal,
but an experimentally more feasible case would be to release the atoms in
a rectangular trap\cite{Box}  i.e. with a flat bottom or on a magnetic mirror\cite{trampoline} so that all the atoms are bouncing
in parabolic trajectories, and scan the optical box across so as to ``skim'' all the trappable atoms from the top,  and slowly move down to capture progressively less energetic  atoms. With this arrangement the box is always at the top, so there is no worry  about unnecessarily stirring atoms using a depopulation beam  and losing them. This will be presented in our future work\cite{Choi2}.

\end{document}